\documentclass[preprint,showpacs,superscriptaddress,amsmath,amssymb]{revtex4}
\usepackage{dcolumn}
\begin{document}

\preprint{RPC}

\title{Partial dynamical symmetries in the $j=9/2$ shell---progress and puzzles}
\author{L.~Zamick}
\affiliation{Department of Physics and Astronomy, Rutgers University,
Piscataway, New Jersey 08854, USA }

\author{P.~Van~Isacker}
\affiliation{Grand Acc\'el\'erateur National d'Ions Lourds,
CEA/DSM--CNRS/IN2P3, BP 55027, F-14076 Caen Cedex 5, France}

\date{\today}
\begin{abstract}
We present analytic proofs of the properties of
solvable states of four particles in the $j=9/2$ shell
which have seniority $v=4$ and angular momentum $I=4$ or 6.
We show in particular that the number of pairs with angular momentum $I$
is equal to one for these states.
\end{abstract}
\maketitle

\section{Statement of the problem}
\label{s_state}
It is known since long that a rotationally-invariant Hamiltonian
describing identical fermions in a single-$j$ shell
has eigenstates with good seniority
for any interaction between the particles as long as $j\leq7/2$.
Proofs of this statement can be found in the books
of de-Shalit and Talmi~\cite{Shalit63} and Talmi~\cite{Talmi93}.
This property of seniority conservation
is no longer valid for all eigenstates in a $j=9/2$ shell
but it turns out that some selected eigenstates
have good seniority for an arbitrary interaction.
More specifically, it was noted in Refs.~\cite{Escuderos06,Zamick07}
that in the techniques used to calculate
coefficients of fractional parentage (CFPs) in Ref.~\cite{Bayman66}
the two $v=4$ states (denoted {\it e.g.} as $|j^4,1,v=4,I\rangle$ and $|j^4,2,v=4,I\rangle$) are degenerate.
Hence the emerging states are arbitrary
and any linear combination of the two
which we may define as
\begin{eqnarray}
|j^4,a,v=4,I\rangle&=&\alpha|j^4,1,v=4,I\rangle+\beta|j^4,2,v=4,I\rangle,
\nonumber\\
|j^4,b,v=4,I\rangle&=&-\beta|j^4,1,v=4,I\rangle+\alpha|j^4,2,v=4,I\rangle,
\label{e_expansion}
\end{eqnarray}
would be equally valid.
If, instead of a pairing interaction,
one uses an arbitrary but seniority-conserving interaction
({\it e.g.}, a $\delta$ interaction)
then the two states become non-degenerate and well defined.
Both eigenstates are independent of the interaction
as long as it conserves seniority.
In addition, one of the linear combinations, say $|j^4,a,v=4,I\rangle$,
has the interesting property that it cannot mix with the $v=2$ state
even if an interaction is used that does not conserve seniority.
Hence this state satisfies the property 
\begin{equation}
M\equiv\langle j^4,v=2,I|\hat V|j^4,a,v=4,I\rangle=0,
\label{e_offd}
\end{equation}
for an arbitrary interaction $\hat V$ in the $j=9/2$ shell.
For notational convenience
the states $|j^4,a,v=4,I\rangle$ and $|j^4,i,v=4,I\rangle$
henceforth shall be denoted in short
as $|j^4v_aI\rangle$ and $|j^4v_iI\rangle$, respectively.

The property~(\ref{e_offd}) was proven numerically in Ref.~\cite{Escuderos06}
for four particles in a $j=9/2$ shell
coupled to total angular momentum $I=4$ or $I=6$.
Subsequently, the states in question were shown to be solvable
({\it i.e.}, to have a simple closed energy expression) in Ref.~\cite{Isacker08}
by means of a symbolic computation in Mathematica.
The purpose of this paper is to provide analytic proofs of these results
based on generic properties of CFPs discussed in the next section.
Parts of these proofs were already given by one of us~\cite{Zamick07}
but are repeated here for completeness.

\section{Relations between coefficients of fractional parentage}
\label{s_cfp}
Our analytic proofs are based on known special properties of CFPs
which are repeated here for completeness.
First, we note the following relation
between $v$-to-$(v+1)$-particle
and $(v+1)$-to-$(v+2)$-particle CFPs:
\begin{eqnarray}
[j^{v+1}(v + 1,\alpha_1J_1)jJ|\}j^{v + 2}v\alpha J]
&=&(-1)^{J+j-J_1}\sqrt{\frac{2(2J_1+1)(v+1)}{(2J+1)(v+2)(2j+1-2v)}}
\nonumber\\
&&\times[j^v(v\alpha J)jJ_1|\}j^{v + 1}v + 1,\alpha_1J_1].
\label{e_cfptal}
\end{eqnarray}
This relation has been derived in the books
of de-Shalit and Talmi~\cite{Shalit63} and Talmi~\cite{Talmi93};
for example, see Eq.~(19.31) of Ref.~\cite{Talmi93}.

Second, we will use the equivalent of the Redmond recursion relation~\cite{Redmond54}
but for CFPs classified by the seniority quantum number $v$
and for which there are no redundancies.
This modified relation is given by~\cite{Zamick06}
\begin{eqnarray}
\lefteqn{(n+1)\sum_{v_s}
[j^n(v_0J_0)jI_s|\}j^{n+1}v_sI_s]\,
[j^n(v_1J_1)jI_s|\}j^{n+1}v_sI_s]}
\nonumber\\
&=&\delta_{J_0,J_1}\delta_{v_0,v_1}+
n(-1)^{J_0+J_1}\sqrt{(2J_0+1)(2J_1+1)}
\sum_{v_2J_2}
\Bigg\{\begin{array}{ccc}
J_2&j&J_1\\ I_s&j&J_0
\end{array}\Bigg\}
\nonumber\\
&&\times [j^{n-1}(v_2J_2)jJ_0|\}j^nv_0J_0]\,
[j^{n-1}(v_2J_2)jJ_1|\}j^nv_1J_1].
\label{e_cfpzam}
\end{eqnarray}
Note that the sum on the left-hand side of this identity
runs over all seniorities $v_s$
but that the total angular momentum $I_s$ is fixed.

\section{Analytic proof}
\label{s_proof}
As noted above, for four identical particles in a $j=9/2$ shell
there are two $v=4$ states with $I=4$ or $I=6$.
They can be written in terms of three-particle states
in the usual way with three-to-four-particle CFPs,
\begin{equation}
|j^4v_iI\rangle=
\sum_{v_3J_3}
[j^3(v_3J_3)jI|\}j^4v_iI]\,
|j^3v_3J_3,j;I\rangle,
\end{equation}
where the state on the right-hand side
results from the coupling of the angular momentum $J_3$ of the first three particles
with the last particle's angular momentum $j$
to total angular momentum $I$.
For $v_i=4$ the intermediate seniority of the first three particles
necessarily must be $v_3=3$.
We now focus on the intermediate state with $J_3=j$.
Given the expansion~(\ref{e_expansion}),
the following relation holds
\begin{eqnarray}
\lefteqn{[j^3(v=3,J=j)jI|\}j^4v_aI]}
\nonumber\\
&&=\alpha[j^3(v=3,J=j)jI|\}j^4v_1I]+
\beta[j^3(v=3,J=j)jI|\}j^4v_2I].
\end{eqnarray}
We can always choose the coefficients $\alpha$ and $\beta$
such that the CFP on the left-hand side vanishes.
In other words, we {\em define} the special states $|j^4v_aI\rangle$ ($I=4,6$) such that
\begin{equation}
[j^3(v=3,J=j)jI|\}j^4v_aI]=0.
\label{e_cfp0a}
\end{equation}
Furthermore, using the proportionality relationship~(\ref{e_cfptal}),
we can deduce the following property:
\begin{equation}
[j^4(v_aI)jJ|\}j^5,v=3,J=j]=0.
\label{e_cfp0b}
\end{equation}
The result~(\ref{e_cfp0b}) will be crucial in the proof
of the property~(\ref{e_offd}) to which we now turn.

In order to prove that the matrix element $M$ of Eq.~(\ref{e_offd})
vanishes for any interaction,
we must show that
\begin{equation}
M(\lambda)=0,
\qquad
{\rm for}\quad \lambda=0,2,4,6,8,
\end{equation}
where $M(\lambda)$ is the matrix element
for a single component $\hat V_\lambda$ of the interaction defined as 
\begin{equation}
\hat V=
\sum_\lambda\nu_\lambda\hat V_\lambda,
\qquad
\nu_\lambda\equiv\langle j^2;\lambda|\hat V|j^2;\lambda\rangle.
\end{equation}
We obtain an expression for $M(\lambda)$ in two steps. 
First, we eliminate one of the four particles
and get an expression in terms of three-particle matrix elements:
\begin{equation}
M(\lambda)=2\sum_{v_3v'_3J_3}
[j^3(v_3J_3)jI|\}j^4,v=2,I]\,
[j^3(v'_3J_3)jI|\}j^4v_aI]\,
\langle j^3v_3J_3|\hat V_\lambda|j^3v'_3J_3\rangle.
\end{equation}
The second CFP in the sum vanishes for $J_3=j$ by construction
and only the terms with $v_3=v'_3=3,J_3\neq j$ survive.
Therefore, the summation may henceforth be considered
as unrestricted in $v_3=v'_3$ and $J_3$.
The three-particle matrix element in turn
can be expressed in terms of a two-to-three-particle CFP,
\begin{equation}
\langle j^3v_3J_3|\hat V_\lambda|j^3v_3J_3\rangle=
3[j^2(\lambda)jJ_3|\}j^3v_3J_3]^2,
\end{equation}
which is obtained in closed form
from the relation~(\ref{e_cfpzam}) for $n=2$,
\begin{equation}
[j^2(\lambda)jJ_3|\}j^3v_3J_3]^2=
{\frac 1 3}+{\frac 2 3}(2\lambda+1)
\Bigg\{\begin{array}{ccc}
J_3&j&\lambda\\ j&j&\lambda
\end{array}\Bigg\}.
\end{equation}
Putting everything together
we obtain for the $\lambda$ component of the interaction
\begin{eqnarray}
M(\lambda)&=&6\sum_{v_3J_3}
[j^3(v_3J_3)jI|\}j^4,v=2,I]\,
[j^3(v_3J_3)jI|\}j^4v_aI]
\nonumber\\
&&\times
\Bigg[{\frac 1 3}+{\frac 2 3}(2\lambda+1)
\Bigg\{\begin{array}{ccc}
J_3&j&\lambda\\ j&j&\lambda
\end{array}\Bigg\}\Bigg].
\end{eqnarray}
The first ``${\frac 1 3}$'' term in the square brackets vanishes
because of orthogonality of the CFPs~\cite{Shalit63,Talmi93}.
The second ``${\frac 2 3}$'' term in the square brackets
can be evaluated for $\lambda=I$
by use of the Redmond relation~(\ref{e_cfpzam}) for $n=4$ which gives
\begin{eqnarray}
M(\lambda=I)&\equiv&\langle j^4,v=2,I|\hat V_{\lambda=I}|j^4v_aI\rangle
\nonumber\\
&=&5\sum_{v_s}
[j^4(v=2,I)jI_s|\}j^5v_s,I_s=j]\,
[j^4(v_aI)jI_s|\}j^5v_s,I_s=j]=0.
\label{e_offd0}
\end{eqnarray}
The sum vanishes because the only term that contributes has $v_s=3$
for which the second CFP is zero according to the property~(\ref{e_cfp0b}).

We can use this fact to prove the result~(\ref{e_offd}),
that is, that the mixing matrix $M$ vanishes
for a general interaction in the $j=9/2$ shell.
Recall that for identical particles there are
five two-body interaction matrix elements in this shell
corresponding to $\lambda=0$, 2, 4, 6, and 8.
Furthermore, there are four seniority-conserving interactions
and one seniority-violating interaction.
(The number of seniority-violating interactions
is one less than the number of $J=j$ states
for three identical particles~\cite{Shalit63,Talmi93}
and there are two three-particle states with $J=9/2$ for $j=9/2$.)
Four independent seniority-conserving interactions are, for example,
a constant interaction, a pairing interaction,
the two-body part of the $\hat J^2$ operator,
and the $\delta$ interaction.
But we have just found a seniority-violating interaction
which does not admix $|j^4,v=2,I\rangle$ and $|j^4v_aI\rangle$,
namely the interaction $\hat V_{\lambda=I}$.
Hence, we can express the five two-body interaction matrix elements
in terms of these five interactions
which will not admix the $|j^4,v=2,I\rangle$ and $|j^4v_aI\rangle$ states.
Indeed all $M(\lambda)$ vanish.

With similar arguments it can be shown
that there is no coupling via the interaction $\hat V_{\lambda=I}$
between the state $|j^4v_aI\rangle$
and the state $|j^4v_bI\rangle$ that is orthogonal to it.
This has yet to be shown for a general interaction.
It has been shown empirically in Ref.~\cite{Escuderos06}
that the special state $|j^4v_aI\rangle$
is an eigenstate for any interaction---seniority conserving or not.
This means that there is no coupling of this state
with the other $v=4$ state via any interaction.
Note that this was not proved in Ref.~\cite{Zamick07}
but that it did emerge from the numerical solution of the equations
given in Ref.~\cite{Isacker08}.

Next we consider the energy of the $|j^4v_aI\rangle$ states.
In Ref.~\cite{Isacker08} closed energy expressions were obtained
with use of Mathematica,
\begin{eqnarray}
E[(9/2)^4v_a,I=4]&=&
\frac{68}{33}\nu_2+\nu_4+\frac{13}{15}\nu_6+\frac{114}{55}\nu_8,
\nonumber\\
E[(9/2)^4v_a,I=6]&=&
\frac{19}{11}\nu_2+\frac{12}{13}\nu_4+\nu_6+\frac{336}{143}\nu_8.
\end{eqnarray}
With minor modifications of what has been done up to now,
we can explain why the coefficient of $\nu_{\lambda=I}$ is one.
This in fact means that the number of pairs with angular momentum $I$
is equal to one for these states.
At this point we keep the discussion for general $j$ and $I$
and {\em assume} that an analytic expression is available for $E[j^4v_aI]$
which is linear in the two-body matrix elements $\nu_\lambda$,
\begin{equation}
E[j^4v_aI]=
\sum_\lambda x_\lambda\nu_\lambda,
\label{e_hyplin}
\end{equation}
where the coefficients $x_\lambda$ depend implicitly on $j$ and $I$.
The application of this relation for each component $\hat V_\lambda$ of the interaction
leads to the identity
$x_\lambda=\langle j^4v_aI|\hat V_\lambda|j^4v_aI\rangle$.
Via an argument analogous to the preceding discussion
this matrix element can be written as
\begin{equation}
x_\lambda=6\sum_{v_3J_3}
[j^3(v_3J_3)jI|\}j^4v_aI]^2
\Bigg[{\frac 1 3}+{\frac 2 3}(2\lambda+1)
\Bigg\{\begin{array}{ccc}
J_3&j&\lambda\\ j&j&\lambda
\end{array}\Bigg\}\Bigg].
\end{equation}
For $\lambda=I$ the sum can be carried out:
\begin{eqnarray}
x_{\lambda=I}&=&
2+4(2\lambda+1)\sum_{v_3J_3}
[j^3(v_3J_3)jI|\}j^4v_aI]^2
\Bigg\{\begin{array}{ccc}
J_3&j&I\\ j&j&I
\end{array}\Bigg\}
\nonumber\\
&=&
2-1+5\sum_{v_s}[j^4(v_aI)jI_s|\}j^5v_s,I_s=j]^2,
\label{e_diag0}
\end{eqnarray}
where use has been made of the normalization property of the CFPs
and the Redmond relation~(\ref{e_cfpzam}) for $n=4$,
in the first and second step, respectively.
The seniority quantum number $v_s$ in the sum
assumes the values 1, 3, and 5.
For $v_s=1$ the CFP in Eq.~(\ref{e_diag0}) vanishes
because one cannot obtain $v=1$
from four particles with $v=4$ coupled to a single particle.
That the CFP vanishes for $v_s=3$ was shown in Eq.~(\ref{e_cfp0b}).
It turns out that the CFP for $v_s=5$ also vanishes.
This follows from consulting tables of CFPs~\cite{Shalit63,Bayman66},
putting in the appropriate coefficients $\alpha$ and $\beta$
to make the CFP~(\ref{e_cfp0b}) vanish,
and subsequently checking that with the same $\alpha$ and $\beta$
the CFP with $v=5$ vanishes as well as.
In other words, we note from the different tables for $j=9/2$
that the following property holds
for arbitrary states $|j^4,1,v=4,I\rangle$ and $|j^4,2,v=4,I\rangle$,
\begin{equation}
\frac{[j^4(1,v=4,I)jJ|\}j^5,v=3,J=j]}{[j^4(1,v=4,I)jJ|\}j^5,v=5,J=j]}=
\frac{[j^4(2,v=4,I)jJ|\}j^5,v=3,J=j]}{[j^4(2,v=4,I)jJ|\}j^5,v=5,J=j]},
\label{e_cfpunk}
\end{equation}
but we have no analytical proof of it.
We conclude that the sum in Eq.~(\ref{e_diag0}) vanishes,
leading to the final result
\begin{equation}
x_{\lambda=I}=\langle j^4v_aI|\hat V_{\lambda=I}|j^4v_aI\rangle=1.
\label{e_diag1}
\end{equation}

These properties can be used to find the energy expressions of the solvable states.
To illustrate the procedure,
we first consider four identical particles in the $j=7/2$ shell
in which case there is at most a single $v=4$ state
for a given total angular momentum $I$
which again we denote as $|j^4v_aI\rangle$.
The results derived above for the $j=9/2$ shell
are equally valid for $j=7/2$.
The property~(\ref{e_offd0}) is trivial
since any interaction is diagonal in seniority in this shell.
The property~(\ref{e_diag1}) follows from the fact
that in the sum in Eq.~(\ref{e_diag0}) we necessarily have $v_s=1$
since five particles in the $j=7/2$ shell are equivalent to three holes
which must have seniority $v=1$ for $J=j$.
The CFP therefore must vanish since a $v=1$ five-particle state
cannot have four of the particles coupled to seniority four.

The result~(\ref{e_diag1}) can be put to good use as follows.
We assume that the energy of the $|j^4v_aI\rangle$ state
can be written as a linear expression~(\ref{e_hyplin})
in the two-body matrix elements $\nu_\lambda$.
The unknown coefficients $x_\lambda$ can be determined
by choosing different interactions
(defined by the two-body matrix elements $\nu_\lambda$)
for which the energy $E[j^4v_aI]$ is known.
Four such interactions are available:
\begin{enumerate}
\item
The pairing interaction which is obtained for
$\nu_0=1$ and $\nu_2=\nu_4=\nu_6=0$
and yields the energy $E[j^4v_aI]=0$.
\item
The constant interaction which is obtained for
$\nu_0=\nu_2=\nu_4=\nu_6=1$
and yields the energy $E[j^4v_aI]=6$.
\item
The two-body part of $\hat J^2$ which is obtained for
$\nu_\lambda=\lambda(\lambda+1)-2j(j+1)$
and yields the energy $E[j^4v_aI]=I(I+1)-4j(j+1)$.
\item
A single $\hat V_{\lambda=I}$ component
which is obtained for $\nu_{\lambda=I}=1$.
According to the preceding discussion the energy is $E[j^4v_aI]=1$.
\end{enumerate}
For $j=7/2$ and $I=2$ or 4 we have thus a system of four linear equations
\begin{eqnarray}
x_0&=&0,
\nonumber\\
x_0+x_2+x_4+x_6&=&6,
\nonumber\\
-\frac{63}{2}x_0-\frac{51}{2}x_2-\frac{23}{2}x_4+\frac{21}{2}x_6&=&I(I+1)-63,
\nonumber\\
x_I&=&1,
\end{eqnarray}
which can be solved for the unknown coefficients $x_\lambda$
to give the expressions
\begin{eqnarray}
E[(7/2)^4,v=4,I=2]&=&
\nu_2+\frac{42}{11}\nu_4+\frac{13}{11}\nu_6,
\nonumber\\
E[(7/2)^4,v=4,I=4]&=&
\frac{7}{3}\nu_2+\nu_4+\frac{8}{3}\nu_6,
\end{eqnarray}
which is also what is obtained
via conventional techniques based on CFPs.
Note that this derivation also constitutes a proof
that the coefficients $x_\lambda$ in the energy expression~(\ref{e_hyplin})
must be rational numbers.

\section{Concluding remark}
\label{s_con}
We have reported on some progress
in the understanding of the peculiar occurrence
of partial seniority symmetry in the $j=9/2$ shell
and have shown it to be the consequence
of general properties of CFPs.
The matter is not fully settled yet
since we still lack an analytic proof of the relation~(\ref{e_cfpunk}).
Also, although we have a simple derivation of energy expressions in the $j=7/2$ shell,
this is not yet the case for the solvable states in the $j=9/2$ shell.

We thank Igal Talmi and Stefan Heinze for helpful discussions.
We thank the INT-Seattle where some of the added work was done.

\end{document}